\documentclass[sigconf]{acmart}

\usepackage{booktabs} 
\usepackage{outlines}
\usepackage{caption}
\usepackage{subcaption}
\usepackage{color}
\usepackage{textcomp}
\usepackage{algpseudocode}
\usepackage{algorithm}%
\usepackage{amssymb}
\usepackage{makecell}
\usepackage{tightheaders2}
\usepackage{tightparagraphs}
\renewcommand\footnotetextcopyrightpermission[1]{} 
\setcopyright{none}

\acmDOI{}

\acmISBN{}

\acmConference[Submitted for review to SIGCOMM]{}
\acmYear{2020}
\copyrightyear{}

\acmPrice{}

\newcommand{\fig}[1]{Figure~\ref{fig:#1}}
\newcommand{\tabl}[1]{Table~\ref{table:#1}}

\newcommand{\sect}[1]{\S~\ref{section:#1}}

\newcommand{\red}{}
\newcommand{\blue}{}

\begin{document}

\title{Constellation: A High Performance Geo-Distributed Middlebox Framework}

\author{Milad Ghaznavi}
\affiliation{%
  \institution{University of Waterloo}
}
\email{eghaznav@uwaterloo.ca}

\author{Ali Jos\'e Mashtizadeh}
\affiliation{%
  \institution{University of Waterloo}
}
\email{mashti@uwaterloo.ca}

\author{Bernard Wong}
\affiliation{%
  \institution{University of Waterloo}
}
\email{bernard@uwaterloo.ca}

\author{Raouf Boutaba}
\affiliation{%
  \institution{University of Waterloo}
}
\email{rboutaba@uwaterloo.ca}

\renewcommand{\shortauthors}{}

\begin{abstract}
Middleboxes are increasingly deployed across geographically distributed data centers. In these scenarios, the WAN latency between different sites can significantly impact the performance of stateful middleboxes.
The deployment of middleboxes across such infrastructures can even become impractical due to the high cost of remote state accesses.

We introduce Constellation, a framework for the geo distributed deployment of
middleboxes. Constellation uses asynchronous replication of specialized state objects to achieve high performance and scalability.
The evaluation of our implementation shows that, compared with the state-of-the-art~\cite{s6}, Constellation improves the throughput by a factor of 96 in wide area networks.
\end{abstract}

\maketitle

\section{Introduction}
\label{section:introduction}

Middleboxes, such as firewalls, load balancers, and intrusion detection systems are pervasive in computer networks \cite{Sherry12asurvey,netflix-eureka,amazon-elb,maglev}.
The network function virtualization vision enables middleboxes to be flexibly deployed across a network and provisions new instances on demand.
Middleboxes may have multiple instances to satisfy traffic demand and
share state across instances to cooperatively process traffic.

Although the instances of a middlebox are typically deployed in the same data center, there has been an increasing demand for deploying middlebox instances across wide area networks~\cite{krononat,barracuda-cloudgen-firewall,nfv-whitepaper,e2}.
This growth stems from the trend towards building multi-data center applications, 
which necessitates global scale network management. 

In many cases, even though the instances
are connected by high latency wide area links,
it is still necessary for them to share state~\cite{distributed-rate-limiting,jaal,multi-path-tcp-ids,fastflex,azure-asymetric-routing}.
Examples include distributed rate limiters that share traffic information to monitor and limit the traffic of multi-data center applications~\cite{distributed-rate-limiting},
intrusion detection systems with instances across a {\red large} ISP network that share statistics to detect attacks~\cite{jaal,multi-path-tcp-ids,fastflex},
and proxies in a content delivery network that actively share their health statuses~\cite{codeen-reliability}.

Existing middlebox frameworks that support state sharing have focused on optimizing for local area network deployments~\cite{s6,chc,opennf,split-merge}. 
Full control over routing allows these frameworks to maintain affinity between traffic flows to middlebox instances. 
This results in fewer remote accesses since per-flow state remains mostly local to an instance~\cite{s6,chc,split-merge}.
However, in wide area networks, traffic can span multiple administrative domains, giving a middlebox framework much less control over routing.
Asymmetric routing and multipath protocols~\cite{mptcp,sctp} compound this issue because a single flow may traverse multiple instances, thus requiring state sharing to process such flows.
The result is more remote accesses to shared state across wide area links, which increases packet latency and reduces middlebox throughput.
These frameworks use synchronous state access for correctness, which is only practical within a local area network as it can add a network roundtrip delay to each packet.

In this paper, we introduce Constellation, a framework for geo-distributed middlebox deployments.
Constellation provides a state management system that is highly scalable and performant even when middlebox instances are deployed across wide area networks.
It separates the middlebox state from its application logic and abstracts shared state using {\em convergent state objects},
which can be independently updated yet still converge.
Transparent to the middlebox application logic, Constellation asynchronously replicates state objects to other middlebox instances.
Replication makes the state local to each instance, and convergence allows a middlebox instance to mutate state with only lightweight coordination.

Asynchronous replication of convergent state enables more flexible load balancing.
Replication subsumes the need for flow-instance affinity, and enables any middlebox instance to process any packet with high performance, as the instance already has the required state.
Replication also provides seamless dynamic scaling since traffic load can be rebalanced among instances without
waiting for state migration.

Standard conflict free replicated data types (CRDTs)~\cite{crdt-survey,crdts} are convergent objects that can be used to represent the shared state for a class of common middleboxes.
However, to support middleboxes such as intrusion detection systems and network monitors~\cite{bloomfilter-survey,cbf-dpi,cbf-hashing,summary-cache,nitro-sketch}, we also develop new CRDTs including a counting bloom filter and count-min sketch.
These CRDTs rely on an {\em ordering} property that is provided by our framework. 
Moreover, state updates in some middleboxes may necessitate violating CRDT properties.
To address this limitation,
we introduce {\em derivative state objects} to support packet processing where the same set of objects are always accessed together.
This is commonly used in network address translators and load balancers.

The properties of convergent state objects offer unique opportunities for us to build an efficient and reliable multicast state replication layer.
Using the idempotence and commutativity properties of state objects, this layer coalesces state updates for more efficient utilization of wide area network bandwidth.
It also provides higher tolerance for straggler instances, as 
they can receive and apply batched updates to reduce bandwidth and processing resources compared with executing uncoalesced operations.

We implemented Constellation using Click~\cite{click}, and evaluated our framework by comparing its performance with S6~\cite{s6}, the state-of-the-art middlebox framework for local area networks.
Our results show that Constellation scales linearly with throughput and experiences no overhead due to network end-to-end latency.
Over wide area networks, Constellation can process $96\times$ the bandwidth of S6, which was not designed to tolerate latency.
In local area networks, Constellation can process up to 11$\times$ the bandwidth of S6,
which comes from eliminating the heavyweight mechanisms that S6 uses to hide remote state accesses (see \sect{evaluation}).
Finally, we show that the complexity of our middleboxes is similar to synchronous approaches when compared to S6.

\section{Background and Motivation}
\label{section:background}

\begin{figure}[t]
    \centering
    \includegraphics[width=0.27\textwidth]{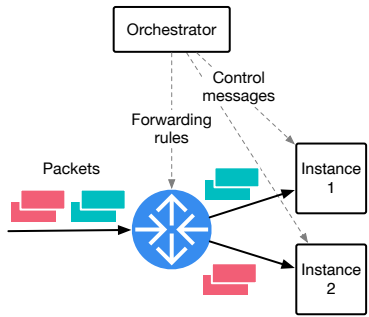}
    \caption{NFV environment:
    \normalfont{An orchestrator distributes the traffic workload among multiple middlebox instances.}}
    \label{fig:nfv-environment}
\end{figure}

\fig{nfv-environment} shows a typical network function virtualization (NFV) environment, where an orchestrator manages middlebox instances deployed on servers and the network connecting these servers.
In response to traffic load, the orchestrator dynamically adds or removes middlebox instances, and installs forwarding rules in the network to redistribute traffic.

The above operations are sufficient to scale stateless middleboxes, e.g., firewalls that pass or block individual packets based on static rules.
However, scaling {\em stateful} middleboxes becomes challenging, since in addition to the aforementioned operations, the middlebox state must be migrated simultaneously with
the new workload distribution~\cite{split-merge,s6,e2}.

\subsection{Middlebox State}
\label{section:middlebox-state}

Stateful middlebox instances maintain {\em dynamic state} regarding traffic flows, which changes how they process packets~\cite{flow-associated,hilti,pico-replication}.
For example, stateful firewalls filter packets based on information collected about flows~\cite{conn-netfliter}.

The middlebox state consists of partitionable and shared state.
The partitionable state is accessed by a single instance, for example
the cache in a web proxy~\cite{squid-memory}.
Shared state can be for a single flow or a collection of flows processed across middlebox instances, and multiple instances query and mutate them.
For example, IDS instances read and update port-counts per external host to detect attacks.

\subsection{Recent Work: State Management for LAN}
\label{section:recent-work}
Existing frameworks that manage shared state are optimized for local area networks and support synchronous accesses to state~\cite{stateless,s6,opennf,split-merge,chc}.
State sharing using this model leads to remote accesses that incur performance cost relative to the network latency.

Two main approaches exist.
One approach separates middlebox state into a remote data store~\cite{stateless,chc}.
Remote state accesses increase packet latency and can reduce throughput by $\sim$60\%~\cite{stateless}, because extra CPU and bandwidth resources are consumed for remote I/Os.
Another approach~\cite{split-merge,opennf,s6} distributes state across middlebox instances.
An instance must query remote instances for non-local state~\cite{s6}.
Frequent remote accesses can significantly degrade performance.
Full control over routing allows the existing frameworks to reduce remote accesses by consistently routing a traffic flow to the same instance so that the flow state remains local to that instance.

Synchronous remote accesses of shared state lead to decreased performance for both approaches.
These frameworks introduce several optimizations to maintain performance of their middlebox implementations.
The state-of-the-art framework, S6~\cite{s6}, masks the overhead of remote accesses using concurrency.
An instance creates a microthread per packet to enable context switching to other packets while waiting on synchronous requests.
However, as we will discuss in \sect{perf-breakdown}, our results show that the overhead of using a microthread per packet halves the maximum throughput of the framework.

Another optimization is to trade consistency for performance.
An instance caches state and performs reads and writes locally~\cite{s6,chc}.
To avoid permanent divergence, cached state copies must be merged periodically.
Doing so na\"ively can result in the consistency anomalies, such as lost updates.

These optimizations complicate the middlebox design.
To achieve acceptable levels of performance and scalability, developers may need to use a combination of these mechanisms.
Reasoning about their correctness is complicated; an incorrect usage can be a source of subtle bugs in middlebox applications.

\subsection{Geo-distributed Middleboxes}
\label{section:middlebox-elastic-scaling}
Middleboxes are increasingly being deployed across wide area networks, e.g., ISP networks, multiple data centers, and content delivery networks.
In such deployments, middlebox instances share state to cooperatively process traffic.

For example, rate limiter instances monitor and limit traffic loads from multiple locations in a content delivery network~\cite{distributed-rate-limiting,codeen-reliability}.
They share their state so that they can limit the global traffic load of multi-data center applications~\cite{distributed-rate-limiting} and control the impact of traffic spreaders~\cite{codeen-reliability}.

IDS instances deployed across an ISP network~\cite{jaal}
share their local statistics to detect intrusions~\cite{jaal,multi-path-tcp-ids,fastflex}.
Using network-wide statistics collected from different network locations is essential to detect
attacks, such as port scanning and denial of service~\cite{heavy-hitter-in-dataplane,fastflex,multi-path-tcp-ids,detect-superspreaders}.
Moreover, multiple NAT instances share the same flow table to translate network addresses across an ISP network~\cite{stateful-nat-asymetric-routing}.

In the CoDeeN peer-to-peer content delivery network~\cite{codeen-reliability}, distributed proxies share their health status.
To handle a cache miss, a proxy redirects a content request to another peer that is healthy based on this information.

\begin{figure}[t]
    \centering
    \includegraphics[width=0.85\columnwidth]{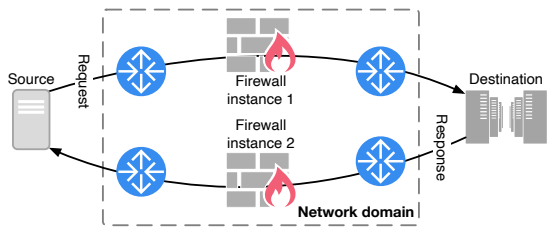}
    \caption{Firewall and asymmetric routing:
    {\normalfont A source creates a connection with a destination.
    The request and response streams of this connection take separate paths and pass through two firewall instances.
    }}
    \label{fig:asymmetric-routing}
\end{figure}

These middlebox deployments face two challenges.
The first challenge is due to characteristics of wide area network traffic. Traffic can span multiple administrative domains with less control over routing. This increases shared state hindering scalability and performance.
Moreover, asymmetric routing and multipath routing~\cite{mptcp,sctp,rexford-hot-potato} undermine the flow-instance affinity.
In asymmetric and multipath routing, a traffic flow,
e.g., sub-flows of a MPTCP session~\cite{mptcp},
may traverse different paths and consequently different middlebox instances.
For a correct operation, the instances must share even per-flow state.

\fig{asymmetric-routing} shows an example asymmetric routing scenario where the traffic of a connection passes through two firewall instances.
A firewall commonly allows connections to initiate only from protected zones, e.g., ``Source'' in \fig{asymmetric-routing}.
State sharing among firewall instances is essential, since the second instance allows the response stream only if
the first instance shares that it has observed the request stream~\cite{stateful-mb,azure-asymetric-routing}.

\begin{table}[tpb]
    \centering
    \begin{tabular}{l l l}
        State Location
        &
        Latency
        &
        Throughput
        \\
        \midrule
        Local Machine
        &
        1--10$\times$~ns~\cite{memory-performance}
        &
        100~M--1~G
        \\
        Remote LAN
        &
        10--100$\times$~$\mu$s~\cite{datacenter-latency,one-way-latency}
        &
        10~k--100~k
        \\
        Remote WAN
        &
        10--100$\times$~ms~\cite{mdcc,geo-state-replication}
        &
        10--100
        \\
    \end{tabular}
    \caption{Time to access state in different locations:
    {\normalfont The throughput of remote state across a wide area network can be as low as 10 to 100 accesses per second.}}
    \label{table:state-access-time}
\end{table}

The second challenge is wide area network latency making state sharing extremely costly.
\tabl{state-access-time} lists the access times to shared state when it is stored locally, or remotely over a local and wide area networks.
Existing frameworks~\cite{chc,stateless,s6}, designed for infrequent state sharing in local area networks, cannot tolerate frequent state sharing over networks with higher latency~\cite{replicated-commit,mdcc,geo-state-replication,multi-datacenter-sdn}.

\section{Design Overview}
\label{section:system-design}

\begin{table*}
\setlength\tabcolsep{1pt}
\centering
    \begin{tabular}{l l l l l l}
        \textbf{Middlebox}
        &
        \textbf{State}
        &
        \textbf{Purpose}
        &
        \textbf{Shared}
        &
        \textbf{Abstract data type}
        &
        \textbf{Size (B)}
        \\
        \midrule
        \makecell[l]{
            IDS/IPS\\\cite{snort,bro,cbf-dpi,flowsize-distribution,nitro-sketch}
        }
        &
        \makecell[l]{
            Session context\\
            Connection context\\
            Bloom filter\\
            Flow size distribution\\
            Port scanning counter\\
        }
        &
        \makecell[l]{
            Session inspection\\
            Connection inspection\\
            String rule matching\\
            Traffic summaries\\
            Port scan detection\\
        }
        &
        \makecell[l]{
            \checkmark\\
            \checkmark\\
            $\times$\\
            \checkmark\\
            \checkmark\\
        }
        &
        \makecell[l]{
            Map\\
            Map\\
            Bloom filter\\
            Count-min sketch\\
            Counters\\
        }
        &
        \makecell[l]{
            96$\times c$\\
            $\sim$250$\times c$\\
            6250\\
            20~k--200~k\\
            112$\times n$\\
        }
        \\
        \midrule
        \makecell[l]{
            Firewall\\
            \cite{npf,pf,real-ip-filter,conn-netfliter,stateful-mb}
        }
        &
        \makecell[l]{
            Flow table\\
            Connection context
        }
        &
        \makecell[l]{
            Stateful firewall/dynamic rules \\
            Connection inspection
        }
        &
        \makecell[l]{
            \checkmark\\
            \checkmark
        }
        &
        \makecell[l]{
            Map\\
            Map
        }
        &
        \makecell[l]{
            32$\times c$\\
            264$\times c$
        }
        \\
        \midrule
        \makecell[l]{
            Network monitor\\
            \cite{tdg-apps,tdg,heavy-hitter-in-dataplane,iceberge}
        }
        &
        \makecell[l]{
            Traffic dispersion graph\\
            Heavy hitters
        }
        &
        \makecell[l]{
            Anomaly detection\\
            Tracking top elephant flows
        }
        &
        \makecell[l]{
            \checkmark\\
            \checkmark
        }
        &
        \makecell[l]{
            Graph\\
            Count-min sketch
        }
        &
        \makecell[l]{
            256$\times n$\\
            -
        }
        \\
        \midrule
        \makecell[l]{
            Load balancer\\
            \cite{katran,maglev,haproxy}
        }
        &
        \makecell[l]{
            Server pool\\
            Server pool usage\\
            Flow table\\
        }
        &
        \makecell[l]{
            Available backend servers\\
            Usage of backend servers\\
            Connection/Session persistence\\
        }
        &
        \makecell[l]{
            $\times$\\
            \checkmark\\
            \checkmark\\
        }
        &
        \makecell[l]{
            -\\
            Vector\\
            Map\\
        }
        &
        \makecell[l]{
            20$\times n$\\
            28$\times n$\\
            28$\times c$\\
        }
        \\
        \midrule
        NAT~\cite{pf,krononat}
        &
        \makecell[l]{
            Available address pool\\
            Flow table\\
        }
        &
        \makecell[l]{
            Tracking available addresses\\
            Address mapping\\
        }
        &
        \makecell[l]{
            \checkmark\\
            \checkmark\\
        }
        &
        \makecell[l]{
            Set\\
            Map\\
        }
        &
        \makecell[l]{
            80$\times c$\\
            74$\times c$\\
        }
        \\
        \midrule
        \makecell[l]{
            Web proxy\\
            \cite{squid,summary-cache}
        }
        &
        \makecell[l]{
            Stored entries\\
            Cache contents\\
            Cache digests\\
        }
        &
        \makecell[l]{
            Metadata of cached contents\\
            Caching in memory or storage\\
            Compact cache summary
        }
        &
        \makecell[l]{
            $\times$\\
            $\times$\\
            \checkmark
        }
        &
        \makecell[l]{
            Map\\
            -\\
            Counting bloom filter
        }
        &
        \makecell[l]{
            104$\times c$\\
            Available storage\\
            20$\times c$
        }
    \end{tabular}
    \caption{Examples of common middleboxes:
    {\normalfont
    A list of common middlebox applications are shown.
    For ``Size (B)'', $c$ and $n$ are respectively the number of connections/sessions and hosts/servers.
    Note that we provide a representative set of state for each middlebox, and the they are not exhaustive. Moreover, for each middlebox application, we list the state of multiple implementations, and a single implementation does not necessarily include all the presented state.
    }}
    \label{table:mbs}
\end{table*}

\tabl{mbs} shows a list of popular middleboxes with a selection of their state.
This list is not exhaustive but provides a representative set of abstract data types used by middleboxes.
To better understand the needs of middleboxes in wide area networks, we start by examining the needs of these popular middleboxes.

\subsection{Study of Common Middleboxes}
\label{section:study}

Our study reveals two key observations.
First, most middleboxes maintain shared state for collecting traffic statistics, resource ownership, and resource usage.
Second, middleboxes mostly operate on relativity small shared state and trade off precision for higher scalability and performance~\cite{jaal}.
A corollary is that most operations on shared state are simple even when the middleboxes are not~\cite{chc,stateless}.

\paragraph{Purpose of sharing state:}
Middleboxes collect statistics about traffic for detection and mitigation purposes.
As shown in \tabl{mbs}, an IDS/IPS track statistics of traffic connections and sessions to detect abnormal or malicious communications.
The instances of a signature based IDS need to share their statistics to detect advanced attacks that can exploit multi-path routing in a wide area network.
These attacks split their signatures across multiple paths to circumvent traditional signature based detection approaches~\cite{multi-path-tcp-ids}.
A stateful firewall inspects the collected statistics to block malicious connections and maintain dynamic rules for outgoing connections.
As mentioned before, the firewall instances may be required to share their state to handle asymmetric flows~\cite{azure-asymetric-routing}.
A network monitor maintains a traffic dispersion graph that embodies the communications between network nodes.
A network wide representation can monitor thousands of hosts to detect large scale attacks~\cite{tdg,gr-ids}.

Middleboxes also track resource ownership and usage for resource management purposes.
As shown in \tabl{mbs}, a load balancer manages backend servers, and distributes load among them based on their usage.
A NAT manages a set of available public addresses and allocates these addresses among network flows.
NAT instances in an ISP network share their state to correctly route asymmetric flows~\cite{stateful-nat-asymetric-routing}.

\paragraph{Shared state implementation:}
Shared state tends to be small, which reduces communication overheads between instances~\cite{multi-path-tcp-ids} and per-packet processing costs.
For the middleboxes shown in \tabl{mbs}, to serve millions of flows, the shared state requires only few 100~MB of the memory.
For example, the most memory intensive middleboxes are web proxies that keep a large cache local to each instance~\cite{squid-memory}.
Advanced proxies share a compact summary of their cached
contents~\cite{summary-cache} to allow redirecting content requests to nearby instances.
This improves the quality of service in serving content requests in a content delivery network.

Although many middleboxes make complex decisions based on shared state, their operations on shared state are simple.
Others have also observed that middleboxes operate on shared state with a simple set of operations~\cite{chc,stateless}.
For example, an IDS collects lightweight packet summaries, but performs complex detection operations locally.
Rate limiter instances
share their observed flow rates~\cite{distributed-rate-limiting} and use probabilistic analysis to shape traffic of multiple data centers.

Middleboxes collect approximate statistics when collecting precise statistics leads to high memory or processing overheads.
They sometimes use compact and approximated statistics for a faster request serving.
As shown in \tabl{mbs}, IDSes and network monitors often use count-min sketches or bloom filters, which are probabilistic data structures, to track top heavy hitters.
Web proxy uses cache digests to quickly check local contents when serving requests.

\subsection{Constellation Design Choices}
\label{section:design-choices}
Our observations lead us to design Constellation.
{\blue Constellation is a geo-replicated middlebox framework that deploy a cluster of middlebox instances distributed across a wide area network.
Constellation provides for the management of shared state across the entire deployment.}

Constellation separates the design of middleboxes into middlebox logic and middlebox state to hide the complexity of state sharing from the middlebox logic.
Transparent to the logic, Constellation asynchronously replicates updates to shared state from each middlebox instance.
Using convergent state, Constellation eliminates most of the complexity of managing asynchrony.
{\blue Our key design choices are as follows.}

\paragraph{Asynchronous state replication:}
Constellation replicates middlebox state to all instances asynchronously.
Each middlebox instance collects and sends its updates of shared state to all other instances in near real time, and they apply these updates to their state locally.
All instances access shared state locally without querying remote instances.

Asynchronous access to local state allows middlebox instances to share state over high latency links of a wide area network.
Replication also supports flexible load distribution and seamless dynamic scaling by subsuming the need for flow-instance affinity.
If a middlebox instance is overwhelmed, traffic can be immediately rerouted to an existing instance that is underutilized.
In removing an excess instance, the orchestrator can reassign traffic load from this instance to another without waiting for state migration.

Storing a replica of shared state requires more memory than that of existing frameworks, but the overhead is not substantial.
As we observed in \sect{study}, many middleboxes operate on lightweight shared state with small memory requirement.
Even larger memory usage does not change the cost of running middleboxes in the cloud, where computation to memory ratios are fixed. Middlebox applications already require substantial compute resources that usually goes hand-in-hand with more than enough memory.

Asynchronous state replication also trades the consistency of state across instances for performance.
For many cases, our state model framework automatically resolves inconsistencies using {\em convergent state objects}.

\paragraph{Convergent state objects:}
Constellation provides a set of state objects to develop the middlebox state.
These objects are guaranteed to be \emph{convergent}, i.e., middlebox instances will observe the same local value for a state object after they receive and perform the same set of state updates applied in other instances.
Convergence eliminates complexities that may arise due to asynchronous state replication assuring developers about the correctness of shared state.

As we discussed in \sect{study}, most middleboxes perform simple operations on shared state.
For these middleboxes, Constellation's builtin convergent state objects can be used to implement their shared state.
For more complex cases, Constellation provides mechanisms for developers to customize state objects to reconcile conflicting state updates.

Asynchronous state replication even when using convergent state objects may introduce some artifacts.
We discuss these artifacts and their impact on middleboxes in \sect{artifacts-discussion}.

\paragraph{Centralized orchestration:}
A logically centralized orchestrator~\cite{onap-whitepaper,e-cord,WL2,contrail}
monitors traffic load, scales the number of instances according to load variations, and distributes the load among the instances.
Constellation does not involve the orchestrator in state replication to avoid creating a potential performance bottleneck.

\section{Constellation Middlebox Framework}
\begin{figure}[t]
    \centering
    \includegraphics[width=\columnwidth]{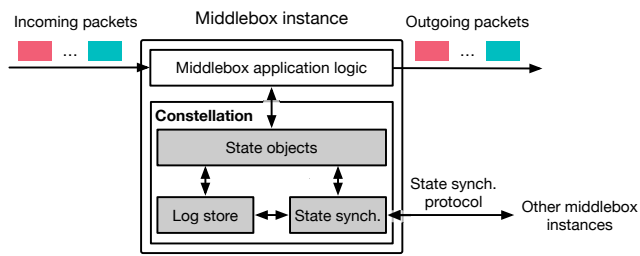}
    \caption{Constellation Middlebox Framework}
    \label{fig:system-design}
\end{figure}
Constellation is designed to allow developers to create geo-distributed middlebox applications with seamless scalability and network latency tolerance.
\fig{system-design} shows a middlebox instance and the main components of the Constellation framework.
Our framework works as follows.

The {\em middlebox logic} registers a set of {\em state objects} that model the middlebox state.
The state objects provide APIs to access and mutate state, which are used by the middlebox logic during its packet processing (discussed in \sect{middlebox-state-representation}).

Constellation internally tracks local state updates by recording them into its {\em log store}.
The \emph{state synchronization} component replicates the recorded state updates to other instances concurrent with packet processing.
All other instances will apply state updates received by the state synchronization component to their local state objects (discussed in \sect{asynchronous-state-replication}).

During adding or removing of middlebox instances, Constellation adjusts the membership of the middlebox cluster while keeping other instances to process traffic.
Constellation selects only one existing instance as a source of the state in bringing a new instance up-to-speed;
other instances experience almost no performance disruption by this membership change (discussed in \sect{dynamic-membership}).

\subsection{State Objects}
\label{section:middlebox-state-representation}

A state object encapsulates a set of variables and operations to access and update their values.
Operations are designed to guarantee that state objects remain convergent, which simplifies reasoning about asynchronous state replication.

To ensure convergence, Constellation uses data structures based on CRDTs~\cite{crdts,crdt-survey} for its state objects.
Two general types of CRDTs exist \cite{crdt-survey}: i) {\em state based} CRDTs where a state update is performed completely local, and the {\em entire} state object is disseminated for state synchronization; ii) an operation based CRDTs where operations are propagated.
Due to the high bandwidth overhead of state based CRDTs, we opt for operation based CRDT.

For convergence, each instance sends its local update operations to other instances, and each downstream instance applies the received operations.
All operations are \emph{idempotent} and \emph{commutative} so instances can safely apply the operations out-of-order and still converge~\cite{crdt-survey}.

We create a library of convergent state objects based on our study of various middleboxes shown in \tabl{mbs}.
The library consists of a collection of abstract data types based on operation-based CRDTs, for instance different flavors of {\tt map}, {\tt set}, {\tt register}, and {\tt counter} objects.
{\red These convergent objects can implement the basic state of common middleboxes.}

{\blue Constellation addresses two limitations of existing CRDTs in implementing the state of middleboxes.
First, Constellation provides atomic updates across multiple objects to address the limitation of CRDTs in supporting multi-object updates.
Second, Constellation develops a {\em convergent} counting bloom filter and count-min sketch which are common in intrusion detection systems, network monitors, and proxies (see \tabl{mbs}). These objects have not been proposed in prior work to the best of our knowledge.
}

\paragraph{\blue Multi object updates:}
Middleboxes need to mutate multiple objects simultaneously,
but this may violate the properties of CRDTs.
This limitation is because convergence is guaranteed for operations on individual CRDT objects~\cite{crdt-survey}.
Reasoning about the convergence of general multi-object operations is complicated,
because the objects may diverge when mutations on multiple objects do not commute.

{\blue
In operating on multiple state objects, there are middleboxes that always update the objects together.
This is common in managing resources where middleboxes update resource usage or ownership when they allocate or release the resources.
For example, a NAT updates an address pool and a flow table object together in processing a new flow, and
a load balancer updates the server pool usage and its flow table together in assigning a new flow to a backend server.
}

{\blue
Operations on multiple state objects do not violate convergence when two conditions hold. First, such a multi-object operation commutes with other operations defined for the objects. Second, the operation is idempotent.
}

{\blue
Constellation supports multi-object operations by defining a derivative object that contains multiple state objects, and performs operations on its state objects simultaneously.
Constellation {\em atomically} replicates this derivative object in a downstream instance to ensure convergence.
{\blue We discuss using derivative state objects to develop a NAT in \sect{nat-implementation}.}}

We also take advantage of the {\em ordering property} of our system to develop a convergent counting bloom filter object and count-min sketch object.
We discuss the ordering property in \sect{asynchronous-state-replication}.

\paragraph{Counting bloom filter:}
A counting bloom filter (CBF) is a memory efficient object for approximate counting.
CBF is used in packet classification, deep packet inspection, and network monitoring \cite{bloomfilter-survey,cbf-dpi,cbf-hashing,summary-cache} where keeping accurate statistics with fine granularity does not scale to traffic load.

A CBF represents a large set of $n$ counters using a smaller vector of $m$ counters. It uses $k$ hash functions to update the counters.
A CBF exposes \texttt{count} and \texttt{value} operations.
The \texttt{count(x)} operation computes $k$ hash values for an operand \texttt{x} (e.g., \texttt{x} can be a five tuple flow identifier).
Each hash value provides an index $0\leq i < m$, and the CBF increments the counter at $k$ computed indices.
On \texttt{value(x)} operation, the CBF computes the same set of $k$ hashes and returns the minimum value among the relevant counters. The returned value is an approximation, since the exact value of \texttt{x}'s counter is less than or equal to this value.

For convergence, 
\texttt{value} and \texttt{count} operations must be commutative and idempotent~\cite{crdt-survey}.
As \texttt{value} does not mutate any counter, we focus on \texttt{count} operation.
Addition commutes, thus \texttt{count} also commutes; however, addition is not idempotent.
Idempotence can be provided using Constellation's ordering feature.
This feature prevents applying a duplicate {\tt count} operation, thus
a local \texttt{count} operation performed in a middlebox instance is only applied once at any other instance across the entire middlebox 
cluster.

\paragraph{Count-min sketch:}
A count-min sketch (CMS) is a probabilistic data structure similar to counting bloom filter and has application in packet classification and deep packet inspections~\cite{nitro-sketch,reversable-sketch}.

A CMS has $k$ arrays of $n$ counters.
A CMS uses $k$ hash functions to update the counters, one hash function per array.
A CMS provides the same set of operations as a counting bloom filter.
To provide convergence, we use the same technique as that of a counting bloom filter.
The idempotence of {\tt CMS} is provided using Constellation's ordering feature.

\subsection{Asynchronous State Replication}
\label{section:asynchronous-state-replication}

An efficient
replication system for convergent state objects has two requirements.
First, we must deliver and apply all operations to all other instances quickly to achieve fast convergence.
Convergent state objects require a one-to-all dissemination of updates and allow commutativity of updates.
Constellation uses multicast to build a replication layer that allows unordered delivery of state updates.
Multicast uses bandwidth efficiently compared to having $O(n^2)$ point to point transports and reduces latency compared to other topologies,
{\blue e.g., forwarding state updates through middlebox instances one by one.}

Second, we must mitigate {\blue straggler} instances that can slow down the replication for the entire middlebox cluster.
Instances may fall behind because of insufficient bandwidth or processing power.
We use the idempotence and commutativity properties of CRDTs to build a congestion control scheme that coalesces state updates to adaptively meet bandwidth and processing constraints of
{\blue stragglers.}
This congestion control can bound how far behind any given instance is from others.
Increasing coalescing can reduce bandwidth requirements, but state updates fall further behind.

\paragraph{Multicast replication:}
During a middlebox operation,
Constellation records state updates into a log store.
For each state object, the log store allocates a queue of {\em log record}s and maintains a sequence number to track these records.
A log record tracks an operation by recording the local order at which it is performed.
Specifically, a log record denotes that an operation was performed on a set of operands at a particular sequence number.
For example, for an IDS's port counter, log record \texttt{(inc,\{22\},7)} shows that a counter was incremented for SSH port 22 at sequence number 7.

The state synchronization component allocates a multicast group per state object.
Via this group,
middlebox instances
share and exchange their local log records of the object.
For convergence, all log records must be delivered and applied to other instances.
Log records are released once delivered.

Constellation runs two threads at the sender and receiver sides to transmit log records.
Send threads use multicast to send local log records to other instances.
Receive threads receive and apply log records, send acknowledgements, and prunes log records delivered to all other instances.

For each state object, the send thread continuously retrieves outstanding log records, i.e., log records that have not yet replicated in other instances, from the queue of this object, creates a {\em state message} from the records, and sends the message to the associated multicast group.
The send thread round-robins between queues belonging to different objects.

A state message carries a list of log records, and an {\em acknowledgement} vector.
Each instance maintains the acknowledgement vector to track log records received from other instances.
This vector contains the highest sequence numbers seen for each instance.

The receive thread uses the acknowledgement vector to track external log records.
Upon receiving a state message, the receive thread applies the operations from the message, and updates the instance's acknowledgement vector.
The receive thread increments a sequence number of the vector when all log records up to and including this sequence number have been received.

Since operations are idempotent and commutative, state objects converge even when applying duplicate or out of order log records~\cite{crdt-survey}.
The receive thread can be also configured to apply log records in order.
This {\em ordering property} provides idempotence for an easier implementation of state objects that are not intrinsically idempotent.
{\blue We used this property to provide idempotence for {\tt count} operation of counting bloom filter and count-min sketch in \sect{middlebox-state-representation}.}

The receive thread also prunes the log store according to received state messages.
For each object, this thread records the last acknowledgement that it has received from other instances.
The smallest acknowledgement shows the latest log record that has been replicated in all instances.
Accordingly, the receive thread prunes all log records up to the smallest acknowledgement.

We provide reliable multicast by retransmitting lost log records due to packet drops.
If a log record has not been acknowledged, the send thread retransmits the record after a timeout based on the maximum round trip time of any instance.
{\blue If a multicast channel is idle, the send thread periodically transmits keep-alive messages containing the latest acknowledgement vector.}

\paragraph{Adaptive bandwidth optimization:}

{\red There are cases when middlebox instances may fall behind in replication due to transient events, such as temporary congestion in the network.}
In these cases, Constellation uses {\em coalesced} log records instead of sending individual log records.
Coalescing can significantly reduce the bandwidth usage of state replication.

The idempotence and commutativity properties of state objects allow Constellation to coalesce related log records, i.e., the records of state operations modifying the same object.
For example, a series of increments and decrements to a single counter can be represented as adding the sum of the operations.
To coalesce related operations, the send thread calls back into to the associated state object.

Constellation detects instances that are falling behind by monitoring the round trip time (RTT) of each instance in the multicast group.
The receive thread measures the minimum and average RTTs for each instance using acknowledgements.
When the average RTT is higher than the minimum RTT by a set threshold, the instance is marked as {\red congested}.

Upon detection, the send thread starts to send coalesced log records using an adaptive {\em lookahead window} based on RTT.
The instance continuously monitors the average RTT to increase or decrease the lookahead window.
The larger the lookahead window, the more coalescing opportunity.

{\blue Constellation adjusts the lookahead window to trade-off the bandwidth reduction from coalescing and the increased synchronization latency from delaying the transmission of log records.}
The lookahead window size is set based on the throughput.
An instance coalesces log records up to a maximum lookahead or when an acknowledgement is received.
If the acknowledgement arrives early, the instance immediately transmits any already coalesced updates, which effectively reduces the lookahead size.

\subsection{Dynamic Scaling}
\label{section:dynamic-membership}

{\blue A scaling event changes the members of multicast groups and consequently impacts state replication.}
For a correct replication during and after this membership change, Constellation ensures three properties:
i) {\em unique identification}: with a new set of members, each active instance remains uniquely identifiable;
ii) {\em membership agreement}: instances agree upon active members so that their send and receive threads can work in harmony; and
iii) {\em convergence}: the new set of members still remain convergent for all state objects.

We assume that the orchestrator is fault tolerant.
If a failure occurs during a scaling event, the orchestrator reliably detects and notifies all instances of the change.

\paragraph{Scale-out event:}
A new instance joins replication groups and copies a snapshot of middlebox state from an existing instance before starting to process traffic.
Adding a new instance is broken down into four steps.

First, the orchestrator deploys a new instance with unique identifier.
To ensure uniqueness, it is sufficient that the orchestrator generates a new identifier or reuses {\blue the identifier of a removed instance.}

Second, the new instance joins the replication groups and starts to record state messages.
It sends a {\tt join} message containing its identifier to the multicast groups to announce that its joining.
Existing instances confirm receiving a \texttt{join} message by adding the instance to the acknowledgement vector
of its state messages.
Upon receiving this confirmation, the new instance starts acknowledging state messages received from existing instances.
It does so by sending empty state messages with an acknowledgement vector.
The new instance retransmits the \texttt{join} message until all existing instances confirm receiving the message.
This ensures the membership agreement property.

Third, the new instance requests an existing instance for a snapshot of the state and metadata (includes the acknowledgement vector of each state object).
The existing instance executes a {\tt fork} system call~\cite{fork} to duplicate its process to take a state snapshot and transmit it to the new instance.
For the snapshot consistency, {\tt fork} is synchronized with packet processing and state synchronization.

Fourth, upon receiving the state snapshot, the new instance applies log records from state messages recorded since the second step.
Lastly, the new instance notifies the orchestrator to redistribute traffic load to it.

Taking snapshots using {\tt fork} is fast, since memory is not immediately copied.
The memory is marked as copy-on-write; the operating system copies memory after the original or child process modifies it.
Constellation further reduces this overhead by using \texttt{madvise}~\cite{madvise}.
Since the duplicated process does not process incoming packets, Constellation tells the operating system to exclude memory pages reserved for receiving incoming packets.
This significantly reduces the pause time of {\tt fork}.

\paragraph{Scale-in event:}
Another benefit of Constellation is that it can scale-in with no state loss and virtually zero packet loss. Removing an excess instance takes four steps.

First, the orchestrator reroutes the traffic load of the excess instance to other instances.
Due to state replication, other instances have the state necessary to process this load.
Second, the orchestrator notifies the excess instance, and this instance waits for some time for remaining inflight traffic to arrive.
Then, the instance drains its outstanding log records to ensure convergence.
Third, the excess instance sends a {\tt leave} message to a multicasting group to announce that it is leaving.
The instance will retry until all instances acknowledge receiving this message, which ensures membership agreement.
Finally, once all other instances have acknowledged the {\tt leave} message, the excess instance notifies the orchestrator to reclaim all resources.

\section{Implementation and Experience}
\label{section:middlebox-applications}

We built Constellation using the Click modular router~\cite{click}.
It consists of 6141~SLOC for the runtime and 2155~SLOC for the middlebox implementations.
We discuss our development experience in using our system compared to exiting frameworks that provide synchronous middlebox state management.
We dive into the implementation of a flow table and a NAT.
Lastly, we discuss the artifacts caused by the use of asynchronous replication.

\paragraph{Convergent flow table:}
A {\em flow table} is used to track network flows and has application in several middleboxes as shown in \tabl{mbs}.
A flow table is a mapping keyed on a hashing of packet headers with values that can be network addresses or some attributes regarding the flows.

A flow table supports \texttt{add} and \texttt{value} operations.
The \texttt{add(k,v)} operation either inserts flow \texttt{k} and value \texttt{v}, or updates the value of flow \texttt{k} with \texttt{v}. The \texttt{value(k)} operation returns a value associated with key \texttt{k}.

For convergence, we focus on \texttt{add} operation, since \texttt{value} does not mutate the state.
The \texttt{add} operation is idempotent but not commutative.
Enforcing a deterministic ordering on concurrent \texttt{add} operations ``artificially'' makes \texttt{add} commutative.
Specifically, a global ordering across the middlebox cluster determines the winner in a race between two \texttt{add} operations modifying the same key \texttt{k}.

The object exposes a callback that allows developers to customize this ordering.
By default, the object uses a numerical comparison, where \texttt{(k,v)} wins against \texttt{(k,v$^\prime$)} only if the binary value of \texttt{v} is greater than that of \texttt{v$^\prime$}.

\subsection{Network Address Translator}
\label{section:nat-implementation}
A NAT bridges two address spaces~\cite{krononat,rfc2993,rfc3022}.
NAT instances share two state objects.
Using a flow table object, a NAT instance maps traffic flows coming from one address space to another address space.
A NAT instance identifies each flow by a unique port number from an available port pool object~\cite{rfc3022}.
Both flow table and port pool are updated together, thus we can use Constellation's derivative state object feature to support multi object operations on these two objects.

In rare incidents,
due to asynchronous local accesses,
NAT instances may concurrently allocate an identical port number for different flows. This violates the NAT's {\em unique port assignment} invariant, and flow translations may collide.

Constellation's convergent flow table enable us to resolve this inconsistency.
We use its callback so that among two collided flows, a flow with larger numerical value of its five tuple wins the race enabling all instances to converge.

\subsection{Artifacts of Asynchronous Replication}
\label{section:artifacts-discussion}
Asynchronous replication may cause middleboxes to experience temporary inconsistencies until instances converge.
We study a number of middleboxes including the ones shown in \tabl{mbs} for their possible artifacts.
There are three categories of artifacts: lag or reduced precision; packet loss; and duplicates and collisions.
In practice, these artifacts are non-issues, as they are rare and are already mitigated by existing protocols or end user applications.
Our design makes the tradeoff of dealing with small artifacts for substantial performance gains on both local area networks and wide area networks.

\paragraph{Lag or reduced precision:}
The most common problem for most middleboxes is that asynchronous replication induces a lag in measurement or reduces the measurement precision.
For example, an IPS may set a threshold for when it blocks traffic and may lag by approximately the round-trip time between instances.
A distributed rate limiter may be imprecise in its ability to set a limit, but for longer flows it can still maintain a tight bounded error.

\paragraph{Packet loss:}
Packet loss issues can arise for stateful firewalls, NATs and load balancers.
{\red This may occur when traffic passes through a different instance while a connection is being established but before state is synchronized between instances.}
For example in \fig{asymmetric-routing}, the second firewall instance may receive the response traffic before its state is synchronized with the first instance.
Since most protocols will retry dropped packets, this artifact will only result in a small increase in latency.

\paragraph{Duplicates and collisions:}
NATs and load balancers may also suffer from collisions or duplicate mappings, if two instances simultaneously generate conflicting mappings.
For example, two NAT instances might reuse the same public IP and port for two connections to the same destination IP and port.
In this scenario we can terminate one connection and have the client to reconnect.
For even very large networks this is exceedingly rare, and disconnects from other issue sources would be orders of magnitude more common.

An alternative approach is to design a NAT with an extra table to allow instances to acquire leases on regions of the public IP and port space.
When a connection arrives the instance would allocate out of one of these pools, thus preventing collisions.
This would require the middlebox to eagerly reserve new ranges when it is running low on the current pool.

\section{Evaluation}
\label{section:evaluation}

We start with a description of our setup and methodology in \sect{setup}, and then we measure the overhead of Constellation framework in \sect{perf-breakdown}.
We measure Constellation's performance during its normal operation and dynamic scaling in \sect{normal-operation} and \sect{dynamic-scaling-eval}, respectively.
Then in \sect{inconsistency-artifacts}, we measure the impact of Constellation's artifacts in our IDPS example. Finally, we discuss the implementation complexity in \sect{complexity}.

\subsection{Experimental Setup and Methodology}
\label{section:setup}
We compare Constellation with S6~\cite{s6} and Sharded.
S6 is the-state-of-art in elastic scaling of middleboxes. We use the publicly available implementation of S6~\cite{s6-github}.
A S6's middlebox application runs as a process that uses DPDK toolkit~\cite{dpdk}.
Sharded is a baseline system used to measure the performance upper bound, as middlebox traffic is sharded with no shared state.
Moreover, we use two middleboxes, a NAT and an IDPS\@. Our implementation of IDPS includes only the port-scan detection/mitigation functionality.

We use a server cluster each equipped with a single Intel D-1540 Xeon with 8~cores and 64~GiB of memory. The servers are connected with an Intel Ethernet Connection X557 10~Gbps NIC to a Supermicro SSE-X3348T switch.
A separate 10~Gbps Mellanox ConnectX-3 NIC connected to a Mellanox switch is used as the \emph{state channel} for state synchronization.
All servers run Ubuntu 18.04.

We use MoonGen~\cite{MoonGen} to generate traffic and measure performance.
Traffic from a generator server is sent through a middlebox instance then back to the generator.
We measure latency and total throughput at the traffic generators.
The packet size in our experiments is 64~B.
MoonGen measures end-to-end latency by sending
timestamped 128~B packets while it simultaneously sends load of 64~B packets.
We also developed a tool to accurately timestamp received packets at microsecond granularity, which allows us to accurately measure throughput changes.
Using this tool, we capture the impact of Constellation's dynamic scaling in \sect{dynamic-scaling-eval}.

Unless stated otherwise, we run 5 second experiments and repeat each experiment 10 times.
The confidence intervals of our results are all within 5\%. As a result, we do not report them in our plots.

\begin{table}
\setlength\tabcolsep{1.5pt}
    \centering
    \begin{tabular}{l l l l l l l}
        &
        Toolkit
        &
        No Op.
        &
        Reference
        &
        Read
        &
        Write
        &
        R+W
        \\
        \midrule
        S6
        &
        $11.80$
        &
        $5.96$
        &
        $3.66$
        &
        $2.52$
        &
        $2.08$
        &
        $1.38$
        \\
        Constellation
        &
        $10.00$
        &
        $9.28$
        &
        N/A
        &
        $9.26$
        &
        $9.20$
        &
        $9.20$
    \end{tabular}
    \caption{
    Throughput of a pass-through middlebox in Mpps:
    \normalfont {\blue ``R+W'' denotes read and write.} S6 runs on DPDK, while Constellation uses DPDK+Click adding overhead to the toolkit baseline.  Reference adds the overhead of finding which instance owns a state object. We measure the read and write costs separately and together. The performance difference between Constellation and S6 is the {\em cost of synchronization}.}
    \label{table:breakdown}
\end{table}

\subsection{Performance Breakdown}
\label{section:perf-breakdown}

Table~\ref{table:breakdown} shows a performance breakdown for a pass through middlebox operating on a counter object.
Using this middlebox, we benchmark S6 and Constellation to breakdown the performance cost of common middlebox operations.
We configure the middlebox to either perform no operation,
or perform a read, a write, or a read and write per packet.

S6 runs directly on DPDK, while Constellation is built using DPDK+Click which reduces baseline throughput by $\sim15\%$.
The no-operation measurement shows the cost from the S6 and Constellation frameworks.
{\blue S6 process each packet in a separate microthread, built using Boost coroutines~\cite{boost-coroutines} to allow an independent microthread to process a packet while another microthread is blocked on a remote state access. Context switching between microthreads results in a loss of $49\%$ of its performance.}

The remaining columns measure the cost of reading and writing shared state.
The reference column measures the time required for S6 to discover which instance owns the key of a flow.
The read and write costs are measured separately and together. S6 slows down by a further 76\% percent over the no-operation column, {\blue excluding the cost of microthreads}.

\subsection{Performance in Normal Operation}
\label{section:normal-operation}
We measure the maximum aggregated throughput and the end-to-end latency of NAT and IDPS.
For wide area experiments, we deploy our NAT instances in a simulated WAN.
Using \texttt{tc}~\cite{tc-command}, we configure the servers running instances to artificially add WAN latency~\cite{mdcc,replicated-commit,multi-datacenter-sdn} to the state channel.
For both our LAN and WAN, we use a traffic load where each NAT instance receives 2000 new flows per second.

\begin{figure}[tpb]
    \centering
    \includegraphics[width=0.9\columnwidth]{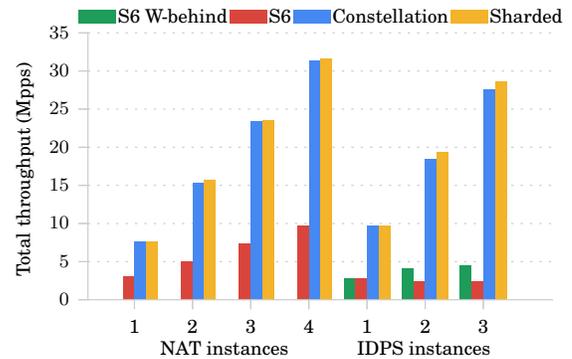}
    
    \caption{Total throughput of middleboxes in LAN:
    \normalfont
    Compared to linear scaling, Constellation is within 2--4\% for NAT and 1--5\% for IDPS.
    }
    \label{fig:throughput}
\end{figure}

\paragraph{Throughput in LAN.}
\fig{throughput} shows the maximum aggregated throughput of the NAT and IDPS instances deployed in our LAN.
For IDPS, we configure S6 in two modes. In the first mode, labeled ``S6,'' the state updates are immediately synchronized.  In the second mode, labeled ``S6 W-behind,'' the remote counters are updated by a 10~ms delay.

As shown for both middleboxes, Constellation's throughput scales linearly with increasing the number of instances, within 2--4\% of the ideal scaling for NAT and within 1--5\% for the IDPS.
For S6, per instance throughput of the NAT drops up to 21\% due to the overhead of state synchronization.
The throughput of IDPS drops for S6 and flattens for ``S6 write-behind'' going from 2 to 3 instances.
In the S6 system, each instance has to query other instances to retrieve the values of their state objects.
IDPS instances pay this overhead once every few packets (i.e., 50\% and 66\% of packets for 2 and 3 IDPS instances), while NAT instances incur this cost once every few flows (i.e., 50\% and 66\% of flows for 2 and 3 IDPS instances).

Compared to S6 for NAT, Constellation improves throughput by 2.5--3.2$\times$ and is within 2--4\% of Sharded's aggregated throughput.
For IDPS, Constellation achieves a 3.4--6.3$\times$ and 3.4--11.2$\times$ higher throughput compared to that of ``S6 write-behind'' and S6, respectively.

\begin{figure}[tpb]
    \centering
    \includegraphics[width=0.9\columnwidth]{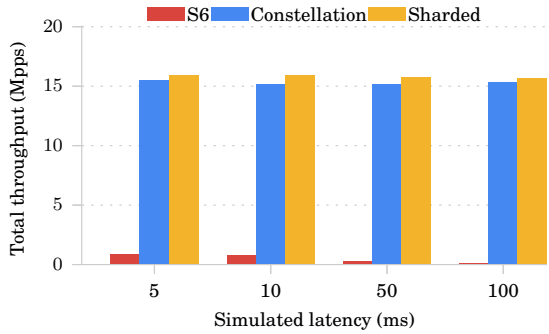}
    \caption{Total throughput of 2 NAT instances in WAN:
    \normalfont Constellation's throughput is largely independent of latency, but synchronous accesses to remote state slow down S6's throughput by $6\times$ to $32\times$ going from 5 to 100~ms latency.}
    \label{fig:wan-throughput}
\end{figure}

\paragraph{Throughput in WAN.}
We evaluate the impact of the state channel with WAN delay on the NAT throughput.
As shown in \fig{wan-throughput}, the aggregated throughput of Constellation's NAT is independent of the WAN delay of the state channel.
However, S6's throughput drops significantly. {\blue Compared to our LAN measurements}, Constellation's WAN throughput is within 2--3\% of its LAN throughput, while S6 becomes 6 to 32$\times$ slower.
Constellation's throughput is 17 to 96$\times$ of S6's and is within 2--5\% of Sharded's.

Constellation accesses the state locally and does not perform immediate state synchronization when a NAT instance writes or queries the state of flows.
This asynchrony allows Constellation's NAT instances to operate at the same throughput level over the WAN as its local area network.
On the other hand, S6's middlebox instances access state stored in a distributed hash table.
Due to state distribution in this hash table, an instance owns a half of the state and must remotely query the other instance to operate on the other half.
The overhead of this synchronous remote access is the root cause of S6's performance drop.

\paragraph{Latency in LAN.}
\tabl{latency} presents the average end-to-end latency of the NAT in our LAN. For a fair comparison, the NAT instances are under S6's sustainable load of 1~Mpps with 2~k new flows per second.
Going from one middlebox instance to two or more instances, both S6 and Constellation enable their state synchronization mechanisms between instances.

\begin{table}
\setlength\tabcolsep{2pt}
    \centering
    \begin{tabular}{l l l l}
        &
        1 instance
        &
        2 instances
        &
        3 instances
        \\
        \midrule
        S6
        &
        $21\pm1\mu$s
        &
        $25\pm1\mu$s
        &
        $26\pm1\mu$s
        \\
        Constellation
        &
        $31\pm1\mu$s
        &
        $44\pm3\mu$s
        &
        $46\pm2\mu$s
        \\
        Sharded
        &
        $31\pm1\mu$s
        &
        $32\pm1\mu$s
        &
        $34\pm2\mu$s
    \end{tabular}
    \caption{NAT average latency:
    \normalfont Constellation's latency remains constant going from 2 to 3 instances. Its latency increase going from 1 to 2 is due to Click's scheduling overhead.}
    \label{table:latency}
\end{table}

As shown in \tabl{latency}, going from 2 to 3 instances, Constellation's latency overhead does not increase.
Compared to Sharded, Constellation adds 12~$\mu$s overhead.
In our implementation, the receive thread of the state synchronization and the middlebox logic run on the same processor core, and we use Click's scheduler~\cite{click} to schedule them.
The latency increase from 1 to 2 instances is due the overhead of Click scheduler~\cite{click}.
S6's latency slightly increases going from 1 to 2 and 3 instances. Its latency is lower than Sharded, since S6's middleboxes run on DPDK, while Sharded's middleboxes uses DPDK+Click which adds Click's overhead to the baseline DPDK {(recall from \sect{perf-breakdown})}.

\begin{figure*}[tpb]
    \begin{subfigure}[l]{\columnwidth}
        \centering
        \includegraphics[width=\textwidth]{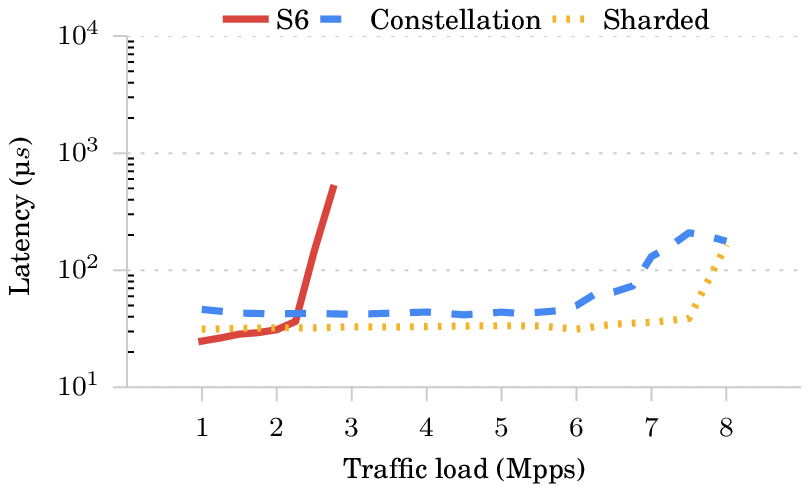}
        \caption{Average end-to-end latency}
        \label{fig:thrlat-avg}
    \end{subfigure}
    \begin{subfigure}[l]{\columnwidth}
        \centering
        \includegraphics[width=\columnwidth]{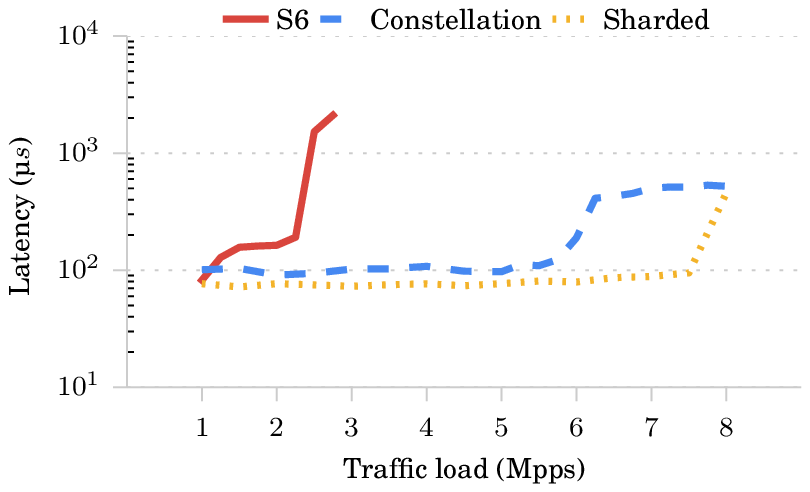}
        \caption{99-percentile end-to-end latency}
        \label{fig:thrlat-99}
    \end{subfigure}
    \caption{End-to-end Latency of the first instance of 2 NAT instances deployed in our LAN:
    {\normalfont Constellation's average and 99-percentile latency remain steady under these traffic loads. When Constellation approaches the saturation point, its latency increases up to 0.2 and 0.5~ms for the average and 99 percentile, respectively.}}
    \label{fig:thrlat}
\end{figure*}

To investigate the latency overhead in more details we report the latency of the first instance of the 2 NAT instances in \fig{thrlat} under different traffic loads.
\fig{thrlat-avg} shows that average latency remains steady for all systems as the traffic load increases until they approach their respective saturation points. Near these points, packets start to be queued, and latency rapidly spikes.
The average latency of Constellation and Sharded remain under 209~$\mu$s, while S6's average latency spikes up to 500~$\mu$s.
As shown in \fig{thrlat-99}, 99-percentile latency has a trend similar to that of the average latency.
Constellation's and Sharded's peak latency values are 451~$\mu$s and 539~$\mu$s, and S6 exhibits a peak latency of up to 2.1~ms.

\subsection{Dynamic Scaling}
\label{section:dynamic-scaling-eval}
\begin{figure}
    \centering
	\includegraphics[width=\columnwidth]{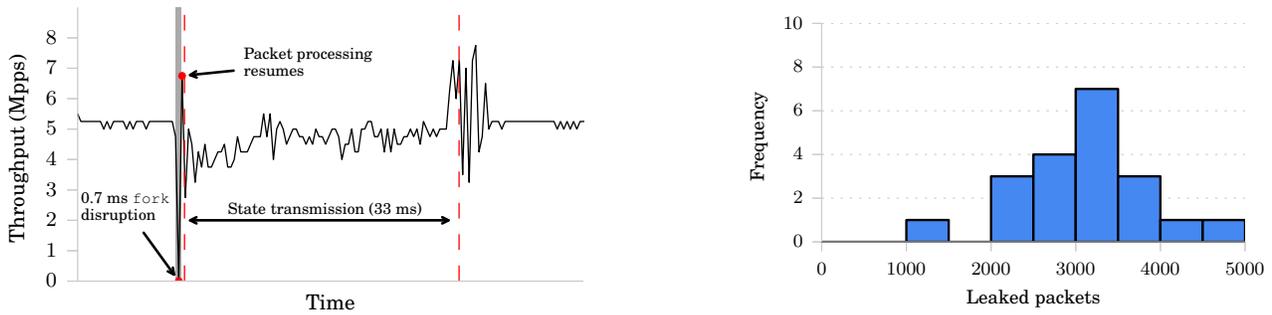}
    \caption{Throughput of the NAT instance transmitting its state in a scale-out event:
    \normalfont The instance experiences a sub-millisecond throughput disruption during \texttt{fork}. The throughput becomes unsteady for few milliseconds during state transmission of 20~k concurrent flows.
    }
    \label{fig:dynamic-scaling}
\end{figure}

We use our cluster of two NAT instances to quantify the performance of Constellation during dynamic scaling.
Each instance is under a 5~Mpps load with 2~k new flows per second.
We use our tool to measure throughput at microsecond scale resolution.
We discuss only the performance of the instance that is involved in state transmission to the new instance.
This instance resides in the same local area network as the new instance.

As shown in \fig{dynamic-scaling}, the first instance does not experience notable throughput degradation. Packet drop is also zero.
Excluding unnecessary memory pages from copy-on-write protection allows \texttt{fork} to complete in only a fraction of a millisecond.
In a separate experiment, not shown here, we measured that \texttt{fork} lasts for 10~ms without this optimization.

Once packet processing resumes, we observe a throughput burst for packets queued during \texttt{fork} pause time.
During state transmission, throughput temporarily fluctuates.
This is due to state updates in processing the first packets of new flows, since they write into copy-on-write memory pages containing the state and incur memory copying overheads.

{\blue
\subsection{Coalescing Benefits}
We measure the bandwidth saving by coalescing state updates of a counting bloom filter and a count-min sketch to evaluate coalescing benefits.
We use network traces from a backbone network~\cite{mawi-dataset} where we use only valid IP packets.

}

\subsection{Inconsistency Artifacts}
\label{section:inconsistency-artifacts}

\begin{figure}[tpb]
    \centering
    \includegraphics[width=0.9\columnwidth]{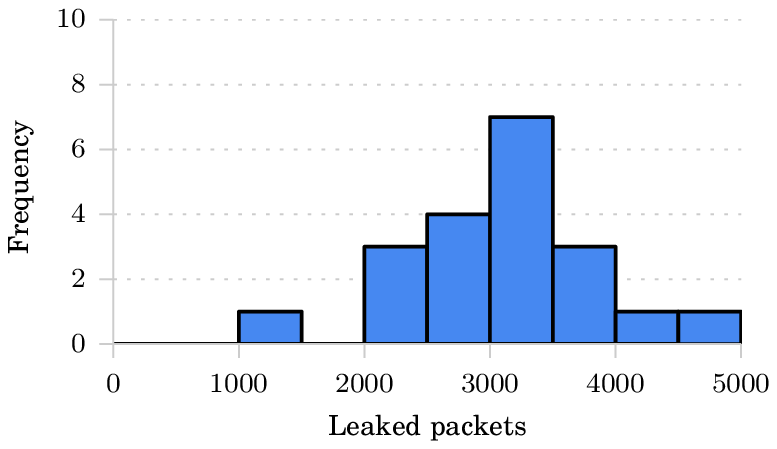}
    \caption{IDPS leaking packets due to asynchronous replication:
    {\normalfont Histogram shows the number of packets leaked beyond the target threshold for Constellation's IDPS.}}
    \label{fig:histogram}
\end{figure}

We measure the impact of Constellation's asynchronous replication in mitigating flooding a port number.
We deploy two IDPS instances in a simulated WAN with 5~ms delay and configure them with a mitigation policy as follows. An instance blocks traffic destined to a port number if the traffic volume passes the threshold of 1024~Mbits.
Two traffic generators flood the instances at
1~Mpps.
For each instance, we measure the number of its \emph{leaked packets}, i.e., the number of packets that pass through an instance in the distributed deployment compared with 
a theoretical centralized IDPS with infinite packet processing that receives the aggregated traffic and filters packets after crossing a given threshold.

\fig{histogram} shows a histogram of the number of leaked packets.
Asynchronous replication delays blocking the attack.
Constellation IDPS reacts to the flood within 5~ms and leaks on average 3.2~k packets.

Delaying an IDPS response by a few milliseconds is a good trade-off as it allows the system to keep up with the throughput demands of high speed networks.
Previous work has shown that IDSes unable to keep up with the traffic can be bypassed to successfully launch attacks~\cite{bib:doscomplex,bib:idselude,bib:attacknids}.

\subsection{Development Complexity}
\label{section:complexity}
Comparing the complexity of different middlebox frameworks is extremely difficult. Using the lines of code, we provide a rough estimation of the complexity.

We compared the code of the NAT implemented in S6 to the one implemented in Constellation as described in \sect{middlebox-applications}.
Both NATs are roughly the same size, with 361 lines of code for Constellation and 283 lines for S6.  We measured S6's code before the source-to-source translator that provides syntactic sugar to simplify the system implementation.
This result illustrates that it is not significantly difficult to build middleboxes with asynchronous replication.

\section{Related Work}
\label{section:related-work}
We have discussed existing frameworks that support state sharing for general middleboxes in \sect{recent-work}.
As mentioned, they are not optimized for wide area networks.
Next, we compare Constellation with two other lines of related work.

\paragraph{Middlebox specific frameworks:}
Some systems are highly specialized for particular middleboxes.
Most of them deploy middlebox instances as shards with no shared state~\cite{krononat,maglev,yoda,beamer}.
Unlike others, vNIDS \cite{vnids}, a microservice based network intrusion detection system, and Yoda \cite{yoda}, an application layer load balancer, share their state in a central data store.

{\blue
\paragraph{Database replication protocols:}
Two phase commit is a common protocol to replicate transactions in distributed databases.
The protocol supports any transaction using synchronous coordination;
however, it does not scale for a geo-distributed replication, since a transaction involves multiple rounds of message passing between a coordinator and replicas in different sites.

Multi datacenter consistency (MDCC)~\cite{mdcc} optimizes performance by involving a coordinator only when transactions conflict.
MDCC also optimizes commutative transactions with domain integrity constraints (e.g., a bank account balance must remain non-negative with concurrent deposits and withdrawals) by involving a coordinator only when concurrent transactions may violate constraints (e.g., the account balance is close to become zero).

Highly available databases relax generality of transactions or consistency guarantees for higher scalability and performance.
Some systems split data into shards and restrict updates to only a single shard.
Eventually consistent databases~\cite{dynamo} allow asynchronous state replication with complex resolution mechanisms to resolve conflicts; however, these mechanisms can cause consistency anomalies.

}

\section{Conclusions}
\label{section:conclusion}

WAN latency can significantly impact the performance of a stateful middlebox whose instances are deployed across a WAN.
We introduced Constellation, a framework for the geo-distributed middleboxes.
Using asynchronous state replication of convergent state objects,
Constellation achieves high performance and scalability.
Our results show that Constellation can improve middlebox performance by almost two orders of magnitude compared to the state-of-the-art~\cite{s6}.

\bibliographystyle{ACM-Reference-Format}
\bibliography{library}

\end{document}